# Evaluating Machine Learning-Driven Intrusion Detection Systems in IoT: Performance and Energy Consumption

Saeid Jamshidi, Kawser Wazed Nafi, Amin Nikanjam, Foutse Khomh

*SWAT, Polytechnique, Montréal, H3T 1J4, Quebec, Canada*

ABSTRACT

In the landscape of network security, the integration of Machine Learning (ML)-based Intrusion Detection System (IDS) represents a significant leap forward, especially in the domain of the Internet of Things (IoT) and Software-Defined Networking (SDN). Such ML-based IDS are crucial for improving security infrastructures, and their importance is increasingly pronounced in IoT systems. However, despite the rapid advancement of ML-based IDS, there remains a gap in understanding their impact on critical performance metrics (e.g., CPU load, energy consumption, and CPU usage) in resource-constrained IoT devices. This becomes especially crucial in scenarios involving real-time cyber threats that challenge IoT devices in a public/private network.

To address this gap, this article presents an empirical study that evaluates the impact of state-of-the-art ML-based IDSs on performance metrics such as CPU usage, energy consumption, and CPU load in the absence and presence of real-time cyber threats, with a specific focus on their deployment at the edge of IoT infrastructures. We also incorporate SDN to evaluate the comparative performance of ML-based IDSs with and without SDN. To do so, we focus on the impact of both SDN's centralized control and dynamic resource management on the performance metrics of an IoT system. Finally, we analyze our findings using statistical analysis using the Analysis of Variance (ANOVA) analysis. Our findings demonstrate that traditional ML-based IDS, when implemented at the edge gateway with and without SDN architecture, significantly affects performance metrics against cyber threats compared to DL-based ones. Also, we observed substantial increases in energy consumption, CPU usage, and CPU load during real-time cyber threat scenarios at the edge, underscoring the resource-intensive nature of these systems. This research fills the existing knowledge void and delivers essential insights into the operational dynamics of ML-based IDS at edge gateway in IoT systems.

## 1. Introduction

The rapid expansion of the Internet of Things (IoT) has ushered in an era where data flows seamlessly across various sectors, driving profound changes in how devices interact [1][2]. This intricate IoT ecosystem, composed of countless devices, sensors, and intelligent nodes, has fundamentally reshaped how we think about device communication, significantly minimizing the need for human involvement [3]. The integration of Software-Defined Networking (SDN) within the IoT landscape represents a significant step forward, creating a unified IoT-SDN framework that offers centralized control, improved network management, and stronger security measures [4][5].

The rapid expansion of IoT, driven by the interconnection of millions of devices via Wireless Sensor Networks (WSNs), presents significant challenges [6]. These challenges stem mainly from these devices' limited memory, power, and battery life, highlighting the need for optimized computing and advanced data analysis techniques [7]. Deploying SDN within this framework aims to overcome these obstacles by offering a streamlined, secure network infrastructure that facilitates effective resource allocation and enhanced threat management.

Given the widespread security vulnerabilities in IoT networks, such as service disruptions and unauthorized access, the importance of Machine Learning (ML)-based Intrusion Detection Systems (IDS) has grown [8]. ML-based IDS are crucial for protecting network integrity due to their ability to adapt dynamically and effectively identify threats [9][10][11].

However, despite advancements in developing ML-based IDS for IoT, several critical gaps remain, as highlighted by Tekin et al. [12]. While previous research has examined ML-based IDS's performance in controlled, static testbed environments, there is a significant gap in understanding how these systems operate under the dynamic conditions of real-time cyber threats, especially when IoT is integrated with SDN. Moreover, while the potential of SDN to significantly enhance resource management in IoT systems is widely acknowledged [13][14][15], there is a lack of empirical evidence on how SDN interacts with ML-based IDS during cyber threats.

In this study, we set two primary objectives designed to deepen our understanding of network performance metrics in IoT. Firstly, we assess the impact of deploying ML-based IDS at edge gateway, mainly focusing on ML-based IDS performance metrics under real-time cyber threats. Secondly, we explore the impact of integrating SDN with our testbed, again at edge gateway, to evaluate its influence on performance metrics under similar cyber threats. The rationale behind incorporating SDN into our testbed is its potential to improve resource management in IoT systems significantly [16][17]. We conduct a comparative analysis of the performance of seven state-of-the-art ML-based IDSs in two distinct setups: firstly, at the edge gateway, and secondly, in a similar setup augmented with SDN integration at the edge gateway, all under real-time cyber threats. This analysis is designed to elucidate the impact of SDN on performance metrics and resource management in IoT systems, especially highlighting how SDN integration can optimize the operational efficiency and resilience of IoT networks against the backdrop of evolving cyber threats. To summarize, this paper makes the following contributions:

- Assessing performance metrics of ML-based IDS in IoT systems under real-time cyber threats: Our investigation revealed the significant impact of seven ML-based IDS on the performance at the edge, specifically measuring CPU usage, CPU load, and energy consumption amidst cyber threats. Utilizing ANOVA, we clarify the operational consequences of deploying these sophisticated IDSs on the edge.

- Evaluating the impact of ML-based IDS at edge integrated with SDN: we evaluated the performance metrics of seven ML-based IDS at the edge gateway system integrated with SDN. Utilizing ANOVA, we clarify the impact of the integrated SDN with IoT on deploying these sophisticated IDS under real-time cyber threats.

- Proposing a plugin-based ML-based IDS test suite: This test suite comes with a group of available datasets and available ML-based IDSs and allows the users to define their own IoT and SDN applications and test their ML-based IDSs and models in terms of detection accuracy and performance metrics. Researchers can efficiently perform comparative analyses for their algorithms and models with other available algorithms and models. The test suite is publicly available (section 8) for researchers and practitioners to reuse.

*Corresponding author
**Principal corresponding author
✉ jamshidi.saeid@polymtl.ca,
kawser.wazed-nafi@polymtl.ca,
amin.nikanjam@polymtl.ca, foutse.khomh@polymtl.ca
(S.J.K.W.N.A.N.F. Khomh)
ORCID(s):

The remainder of this paper is organized as follows: Section 2 discusses the review of our research literature. Section 3 discusses the necessary background knowledge. In Section 4, we describe the experimental design, the Research Questions (RQs), and the metrics of the experiments. Section 5 explains our results and findings. Section 8 discusses threats to the validity of our study. Finally, Section 9 concludes the paper and outlines future work.

## 2. Related Works

Understanding the performance trade-offs of ML-based IDS in IoT, especially in resource-constrained edge gateways, remains an open challenge. While numerous studies, as mentioned in the previous section, have focused on detection accuracy, limited research has analyzed their real-time computational impact. In particular, there is a significant gap in understanding how ML-based IDS operate under real-time cyber threats, especially when integrated with SDN. This section reviews prior works on ML-based IDS in IoT and SDN, examining their strengths and limitations and focusing on ML models and energy consumption concerns.

### 2.1. IoT Intrusion Detection

Alsulami et al. [18] proposed a new ML model to identify and categorize network activity in IoT systems. Their research aimed to classify network traffic into distinct categories, including normal behavior and various types of attacks ( e.g., Mirai, Denial-of-Service (DoS), Scan, and Man-in-the-Middle (MITM)). The study involved testing several supervised learning models on the customized IoTID20 dataset, including Spiking Neural Networks (SNNs), DT, Boosting Trees (BT), Support Vector Machines (SVM), and KNN. These models, enhanced through deep feature engineering, effectively identified and classified network anomalies.

Mukherjee et al. [19] conducted an in-depth investigation into the predictive capabilities of supervised learning models (e.g., Logistic Regression (LR), Naïve Bayes (NB), DT, RF, and Artificial Neural Network (ANN)) for anomaly detection. Their study utilized a dataset comprising 350,000 data points. The research compared these models against established state-of-the-art techniques, including BIRCH clustering and K-Means, and evaluated their performance in different scenarios. This included an analysis using the complete dataset and a separate evaluation after removing binary data points in the 'value' feature. The models demonstrated high precision in both scenarios, underscoring their efficacy in practical anomaly forecasting and enhancing security measures against potential risks.

Elnakib et al. [20] proposed the Enhanced Intrusion Detection Deep Learning Multi-class Classification Model (EIDM), a sophisticated Deep Learning (DL) model designed to enhance security in the IoT context. This model is adept at accurately categorizing 15 distinct traffic characteristics, encompassing a range of 14 discrete attack types. The performance of EIDM was evaluated against four other contemporary models, focusing on classification accuracy and efficiency. The increased precision of EIDM highlights its promise as a powerful solution for safeguarding IoT networks against a wide range of attacks.

Douiba et al. [21] proposed an innovative IDS to enhance IoT device security. Their approach utilized gradient boosting and DT in the Catboost framework. The model's performance was rigorously assessed on several datasets, including NSL-KDD, IoT-23, BoT-IoT, and Edge-IIoT, with optimization achieved through GPU acceleration. The IDS distinguished itself with its ability to detect anomalies in real-time and its computing efficiency, demonstrating high accuracy, recall, and precision metrics, around 99.9% on a record detection and computation time.

Kasongo et al. [22] presented a research endeavor in which they proposed a Feed-Forward Deep Neural Network (FFDNN) IDS, enhanced by the inclusion of a Wrapper Feature Extraction Unit (WFEU) utilizing the Extra Trees algorithm. The WFEU-FFDNN was evaluated for its performance on several datasets, including UNSW-NB15 and AWID, and compared with traditional ML methods. The system demonstrated high classification accuracies in binary and multiclass classifications across these datasets, significantly outperforming in scenarios involving the AWID dataset. The enhanced precision of the WFEU-FFDNN model emphasizes its efficacy in real-time anomaly detection and computing efficiency.

In addition to all of the works stated above, Verma et al. [23] examined ML algorithms in the context of augmenting security measures in the IoT. The researchers compared classifiers using benchmark datasets (e.g., CIDDS-001, UNSW-NB15, and NSL-KDD). This analysis was supported by statistical tests, namely the Friedman and Nemenyi tests. The researchers also evaluated the reaction times on the Raspberry Pi platform, showcasing the adaptability and efficiency of the classifiers in IoT scenarios, hence emphasizing their practical relevance.

Otoum et al. [24] presented a scholarly investigation in which they propose a DL-powered intrusion detection system (DL-based IDS) to effectively address challenges associated with feature learning and dataset management. The DL-based IDS developed by the researchers integrates the Spider Monkey Optimization(SMO) algorithm with the stacked-deep polynomial network (SDPN) to enhance threat identification. The system can detect various abnormalities, including DoS, User to Root attacks (U2R), probing, and Root-to-local attacks (R2L). The DL-based IDS was evaluated using the NSL-KDD dataset and exhibited outstanding performance metrics, showcasing its efficacy in various aspects of threat detection.

Gaber et al. [25] highlight securing IoT systems, especially in complex environments ( e.g., smart cities). The authors introduced a feature selection methodology that combines constant removal and recursive feature elimination strategies. They utilized a DT classifier with a subset of 8 characteristics, assessed on the AWID dataset using various ML classifiers. In contrast to existing methods, their approach exhibited exceptional performance, achieving high accuracy, precision, and F1 score rates. These results underscore the potential of their methodology in the domain of IoT-IDS.

Sachdeva et al. [26] investigate the issue of fortifying cybersecurity in IoT networks to mitigate the impact of distributed denial-of-service (DDoS) attacks. The authors put out an innovative approach for data pre-processing, which involves the integration of ML and DL classifiers. The class imbalances in the BOT-IoT and TON-IoT datasets from UNSW Australia are mitigated using several Synthetic Minority Oversampling Technique (SMOTE) variants. The hybrid methodology employed in this study, which integrates many algorithms, demonstrates the promising prospects for efficient detection of DDoS attacks in IoT networks.

The related works discussed above show that the most ML-based IDS developed and re-used by researchers are DT, KNN, RF, LSTM, CNN, and a hybrid model of CNN and LSTM. In addition, EIDM is the most recent work that has overcome the limitations of the previous ML models. That is why we proceed with all these six ML-based IDS to carry out our study in this paper.

### 2.2. Energy consumption in IDS

Only a tiny amount of research has been done so far to determine the energy consumption in IDS. Among them, Tekin et al. [12] investigated the topic of IDS in the context of the IoT, with a specific focus on the energy consumption aspect in devices with limitations. The authors assessed various ML paradigms in the context of cloud computing, edge computing, and IoT devices. They specifically emphasize the promising capabilities of TinyML for microcontroller units (MCUs). DT algorithm demonstrates in terms of training, inference, and power efficiency. Although Naive Bayes (NB) has superior training speed, it exhibits a minor accuracy trade-off requirements of the KNN algorithm increase proportionally with the quantity of the dataset, hence diminishing its suitability for deployment in IoT systems. Both DT and RF exhibit low power consumption and high accuracy. However, it is essential to consider that RF's longer execution time represents a trade-off. The research findings also elucidate the advantages and constraints of cloud-based ML, underscoring the significance of algorithm choice in practical implementations.

Nimmy et al. [27] utilize the energy consumption patterns of IoT devices to identify irregularities in smart home environments. They developed a prototype of a smart camera based on Raspberry Pi to gather power traces during regular operations and simulated DDoS attacks. This approach emphasizes the importance of energy consumption as a crucial indicator of aberrant behaviors. The deep feed-forward neural network used in their study demon- strates exceptional performance in identifying anomalies, as evidenced by rigorous evaluations of ML models. This indicates its potential to enhance the security of smart homes significantly.

### 2.3. IoT Intrusion Detection in SDN

Chaganti et al. [28] present a sophisticated IDS for IoT networks. This system leverages SDN and specifically emphasizes the utilization of DL techniques. The research is for its utilization of LSTM networks, a Recurrent Neural Network (RNN) type renowned for its efficacy in handling time series data, which is critical in detecting network threats. The authors' principal contribution is utilizing an LSTM model, which they employ to discern network attacks. To evaluate the efficacy of their approach, the authors conduct a comparative analysis with alternative architectures(e.g., SVM). The experimental findings present solid evidence that highlights the improved efficacy of the LSTM model in accurately categorizing various network attacks. The LSTM model demonstrated exceptional accuracy and efficiency in detecting attack patterns, surpassing conventional ML models in precision and recall metrics.

M. M. Isa et al. [29] present the DAERF model in their research, an innovative IDS for SDN. This model combines

a Deep Autoencoder (DAE) with an RF algorithm, creating a unique approach. The DAE excels in feature extraction and data dimensionality reduction. At the same time, the RF approach, known for using an ensemble of DTs, shows significant accuracy and robustness in classification tasks. The DAERF model was evaluated in a simulated SDN using commonly used datasets, demonstrating a high efficacy level. The integration of DL and ML in the DAERF model represents a novel approach that effectively identifies and categorizes network intrusions, enhancing the security of SDN systems and ensuring their capability to handle real-time applications with scalability and adaptability.

Phan The Duy et al. [30] presented 'FoolYE,' an innovative IDS designed specifically for SDN systems. The system combines cyber deception techniques with Moving Target Defense (MTD) methodologies. The core of this methodology lies in its ability to create a dynamic and misleading network environment, making it challenging for malicious actors to identify and exploit genuine resources. A key innovation is deep transfer learning-based IDS, which employs advanced DL models (e.g., ResNet50 and DenseNet161), originally designed for image recognition. These models have been adapted using deep transfer learning techniques to analyze network traffic for ML-based IDS, demonstrating the versatility and efficacy of DL in cybersecurity. The study involved experiments in simulated SDN systems, where the performance of the IDS was thoroughly examined, showing its high capability in accurately detecting a wide range of network intrusions.

Despite advancements in ML-based IDS for IoT, a significant gap remains in understanding their real-time computational impact, especially in energy consumption, CPU load, and CPU usage at the edge gateway. This gap is further compounded by the lack of empirical studies evaluating the effectiveness and efficiency of ML-based IDS in real-world, resource-constrained edge gateway, especially when integrated with SDN during cyber threats. To address these shortcomings, our study provides a comprehensive empirical analysis of ML-based IDS, focusing on their performance trade-offs in SDN-enabled and non-SDN edge gateways. Specifically, we assess how different ML-based IDS models impact system resources under real-time cyber threats, offering critical insights into their feasibility for deployment in IoT networks.

## 3. Background

This section dives into the underlying premise of the research's baselines.

**Decision Tree (DT):** In the field of IDS, DT is a key ML method for analyzing network data. They use trees, e.g., models, to break down network features into binary decisions, evaluating network attributes at each node to identify effective splits. This creates a rule-based hierarchy that excels at spotting differences between normal and suspicious network activities. DTs are valued for their clarity and ease of interpretation, playing a vital in improving cybersecurity by identifying unusual or unauthorized actions [31] [32].

**Random Forest (RF):** The algorithm is highly valued in IDS for its precision in classifying network data. Utilizing RF, an ML algorithm, it creates a group of DT to assess various network attributes, effectively distinguishing between normal and malicious activities. RF excels in managing large datasets, balancing IDS data disparities, and minimizing overfitting, making IoT and network security crucial. It achieves accurate detection of unusual network behaviors [33] [34].

**K-Nearest Neighbor (KNN):** The KNN algorithm is a key IDS tool known for its effective similarity-based classification. It compares network traffic with existing labeled data using distance metrics to classify new instances, with 'k' indicating the number of neighbors considered. This method is crucial for identifying normal versus abnormal network activities, offering a simple yet versatile solution for real-time IDS. KNN excels in both binary and multiclass problems, providing quick, reliable categorizations crucial for responding to threats in dynamic networks [35] [36] [37].

**Long short-term memory (LSTM):** LSTM networks, a type of recurrent neural network, are highly effective in analyzing sequential data for IDS. Their unique memory cells excel at identifying complex patterns in network traffic, making them adept at spotting advanced threats that traditional methods may miss. LSTMs are especially valuable for maintaining context over data sequences, which is crucial for distinguishing between normal and malicious network activities. Their application in IDS significantly boosts cybersecurity, especially in dynamic and IoT environments, by adapting to new threats and efficiently handling varying data lengths, offering a robust solution to modern cybersecurity challenges [38] [39].

**Convolutional Neural Network(CNN):** CNNs provide a resilient DL methodology for IDS. CNNs are widely recognized for their ability to independently acquire hierarchical features from network traffic. This is achieved through convolutional, pooling, and fully connected layers, which enable the discernment of spatial patterns in the traffic data. This capacity facilitates the recognition of both well-established and new threats. CNN in IDS is considered crucial in enhancing cybersecurity defenses against a wide range of cyber threats due to their capacity to scale effectively and efficiently handle real-time data [40] [41].

**Hybrid model of LSTM and CNN:** The integration of LSTM and CNN models into IDS significantly boosts network security by combining the spatial analysis capabilities of CNNs with the temporal pattern recognition of LSTMs. This hybrid approach detects complex cyber threats by analyzing network traffic data in both spatial and temporal dimensions. CNNs effectively identify security breaches through local pattern recognition, while LSTMs track the sequence of network events over time, offering a detailed understanding of potential threats. This fusion results in more accurate and efficient detection of sophisticated, multi-stage attacks, reducing false positives and adapting to new threats, thereby enhancing overall anomaly detection and maintaining network integrity without excessive alerts [42] [43].

**EIDM:** The EIDM is a cutting-edge IDS approach expertly handling a wide range of network events. Its design combines convolutional and dense layers to tackle the challenges of class diversity and data imbalance. The model begins with a 120-node dense layer, followed by an 80-neuron convolutional layer with a kernel size of 20 to better distinguish between similar network activities. It also features a Maxpooling layer for enhanced feature extraction and a dropout layer to avoid overfitting. EIDM can classify 15 network behaviors through six dense layers, using 'relu' activation and SGD and Adam optimizers for optimal accuracy and efficiency. According to [20], EIDM's unique structure and optimization techniques make it a standout solution for improving network IDS.

## 4. Study design

This section describes our methodology to evaluate the impact of specific ML-based IDSs using selected performance metrics. We first mention our Research Questions (RQs), followed by an explanation of the experimental design and the metrics used to evaluate the impact of the ML-based IDS.

### 4.1. Research questions(RQs)

Our research aims to address the following RQs:

- **RQ1: How do ML-based IDSs impact CPU usage, CPU load, and energy consumption at the edge gateway without SDN during real-time cyber threats?**
  This RQ examines the impact of ML-based IDSs on crucial performance metrics, specifically CPU usage, CPU load, and energy consumption, at edge gateway not integrated with SDN. It focuses on analyzing the performance of seven state-of-the-art ML-based IDSs and their impacts on these key metrics in the face of diverse cyber threats.

- **RQ2: What are the differences in CPU usage, CPU load, and energy consumption impacts of ML-based IDS at the edge gateway with SDN integration during real-time cyber threats?**
  This RQ explores how ML-based IDSs influence CPU usage, CPU load, and energy consumption at the edge gateway integrated with SDN. It involves analyzing the impacts of various ML-based IDSs on these essential performance metrics under various cyber threats.

### 4.2. DataSet

In our study, we used the CICIDS2017 data set [44], a highly regarded resource organized by the Canadian Institute for Cybersecurity. This dataset is recognized as one of the gold standards in cybersecurity research, capturing a broad spectrum of benign network activities and the latest cyberattacks [45]. CICIDS2017 is designed to simulate real-world network environments, making it an essential resource for researchers to test and validate advanced IDS thoroughly. The breadth and diversity of the asset highlight its importance, making it necessary for those aiming to strengthen network security paradigms.

**Table 1**
Distribution of labeled IoT-SDN attacks in the dataset

| IoT Attack Labels | No of labeled entries |
|---|---|
| BENIGN | 2271320 |
| DoS Hulk | 230124 |
| Port Scan | 158804 |
| DDoS | 128025 |
| DoS GoldenEye | 10293 |
| FTP-Patator | 7935 |
| SSH-Patator | 5897 |
| DoS slowloris | 5796 |
| DoS Slowhttptest | 5499 |
| Bot | 1956 |
| Web Attack & Brute Force | 1507 |
| Web Attack & XSS | 652 |
| Infiltration | 36 |
| Web Attack & SQL Injection | 21 |
| Heartbleed | 11 |

### 4.3. The ML-based IDS

Numerous ML-based IDS have been developed by researchers [12] [22] [25] [46]. However, we had a significant challenge in reviewing these publications and selecting some for our study. Most did not make their solutions' applications or source code publicly available. This lack of transparency hinders the ability to experiment with these works in real IoT devices. This omission complicates, and may even prevent, the objective comparison of the proposed solutions. Consequently, to initiate our study, it became necessary to independently implement all ML-based IDS that have been previously utilized, except the ML-based IDS proposed by [20], which shared their code ML-based IDS available to researchers. In this section, we explore the implementation process of seven ML-based IDSs that we have developed: DT, KNN, RF, LSTM, CNN, and a hybrid model of LSTM and CNN. Table 3 presents a comparative analysis of the performance metrics of ML-based IDS.

#### 4.3.1. DT, KNN, RF

We have developed and deployed DT-based IDS, RF-based IDS, and KNN-based IDS [47], each specifically designed to improve security policy. The foundation of these models is a preprocessing technique applied to the selected CICIDS 2017 dataset. The dataset features various simulated cyber-attack scenarios alongside standard traffic data. It encompasses multiple numerical attributes, including but not limited to packet sizes, flow durations, and bytes per flow, which are critical for analyzing network behavior and detecting anomalies. We applied min-max normalization as our initial preprocessing step to ensure uniformity across these diverse numerical attributes and mitigate scale discrepancies. Missing values were imputed to preserve the integrity of the data. The LabelEncoder[48] was utilized to convert labels into a format suitable for ML techniques. An essential aspect of our methodology is to divide the selected dataset into training and testing subsets. For the first RQ, we adopted 80% training and 20% testing, aligning with standard practices in ML model development. This adjustment was made to accommodate the different requirements of each research phase. As shown in Table 1, the dataset has five classes (Benign, DDoS, DoS, Brute force, and Port scan) with significantly more entries than the remaining ten classes, which contain fewer samples. SMOTE [49] with auto-sampling was employed to address the class imbalance issue in the dataset. This technique effectively augmented the representation of underrepresented classes, leading to a more balanced dataset for training purposes.

**Table 6**
Comparison of structure and accuracy of different Neural Network models in IDS for IoT-SDN network

| Dataset | CICIDS2017 | CICIDS2017 | CICIDS2017 | CICIDS2017 |
|---|---|---|---|---|
| Categories | 15 | 15 | 15 | 15 |
| Model | LSTM | LSTM+CNN | CNN | EIDM |
| Layers | 10 | 11 | 8 | 12 |
| Parameters | 56386 | 12795 | 3497 | 48735 |
| Structure details | Dense (64) Dense (128) LSTM (128) LSTM (256) Dense (128) Dense (48) Dense (15) | Dense (64) Conv1D (64, 10) Conv1D (64, 10) MaxPooling1D (2) LSTM (128) LSTM (64) Dense (64) Dense (15) | Conv1D (16,30) Conv1D (16,30) MaxPooling1D (2) Flatten() Dense (32) Dense (15) | Dense (120) Conv1D(80, 20) MaxPooling1D (2) Dense (120) Dense (100) Dense (80) Dense (60) Dense (60) Dense (40) Dense (15) |
| Training Accuracy (%) | 97.72% | 98.77% | 97.92% | 99.57% |
| Testing Accuracy (%) | 93.86% | 95.75% | 94.74% | 99.56% |

*4.3.2. CNN*

In our research, we deployed a CNN-based IDS tailored for our experimental testbed. The configuration details of the CNN model, including its layers, parameters, and architecture specifics, are outlined in Table 2.

*4.3.3. LSTM*

In our investigation, we implemented an LSTM-based IDS specifically for our testbeds. The detailed architecture and parameters of the LSTM model, crucial for its operation in our IDS, are thoroughly presented in Table 2.

*4.3.4. Hybrid model of LSTM and CNN*

In our exploration, we implemented a hybrid LSTM and CNN architectures model to create an advanced IDS tailored to our experimental setup. This architecture has already been tested in various scenarios [50][51][43]. The intricate configuration of this hybrid LSTM and CNN model, which leverages the strengths of both LSTM and CNN to enhance detection capabilities, is detailed in Table 2.

The goal of using the hybridization of LSTM and CNN is twofold. First, CNN can drop the non-impactful features and select only the impactful ones (feature engineering). At the same time, it helps to learn the features in a Spatial Hierarchical manner [52]. Second, from our dataset, we got 77 features. As it is unknown which features are impactful from the given features, we applied a 2 1-dimensional CNN layer followed by a max-pooling layer to find the impactful features by learning the 10 nearby features together (kernel size 10). This helps us to create new feature representations where the impactful ones are sustained. Later, we fed these newly derived features directly to 2 LSTM layers. This step helps to learn the spatial and temporal features from CNN, resulting in feature representations presented in context and awarded. Finally, we applied 2 Dense layers to regress the feature representations generated from previous CNN and LSTM layers into 15 classes. This process helps us learn the input features more deeply and increase the classification accuracy.

**4.4. Experimental Design**

To address RQ1, we designed a testbed incorporating two Raspberry Pi 4 Model B units as edge gateways. Each unit is equipped with 8GB of RAM and a 1.5GHz 64-bit quad-core CPU, providing a realistic environment for evaluating the computational impact of ML-based IDS at the edge gateway. Our study evaluates the performance of seven ML-based IDS models: DT, KNN, RF, LSTM, CNN, EIDM, and a hybrid of LSTM and CNN model, selected for their established effectiveness in cybersecurity. We conducted controlled experiments in IoT-edge networks to assess these IDS models, simulating a range of cyber threats(e.g., BENIGN, DDoS, DoS, Brute force attacks, and the Port scan) using Kali Linux [53]. These experiments enabled us to analyze the IDS models' impact on critical performance metrics, specifically CPU usage, CPU load, and energy consumption.

**Table 3**
Performance Comparison of ML-based IDS

|           | DT     | KNN    | RF     | LSTM   | LSTM+CNN | CNN    |
|-----------|--------|--------|--------|--------|----------|--------|
| **Accuracy**  | 0.9985 | 0.9967 | 0.9981 | 0.9386 | 0.9575   | 0.9474 |
| **Precision** | 0.9985 | 0.9966 | 0.9980 | 0.9771 | 0.9877   | 0.9792 |
| **Recall**    | 0.9985 | 0.9967 | 0.9981 | 0.9524 | 0.9645   | 0.9611 |
| **F1-Score**  | 0.9985 | 0.9966 | 0.9980 | 0.9646 | 0.9760   | 0.9701 |

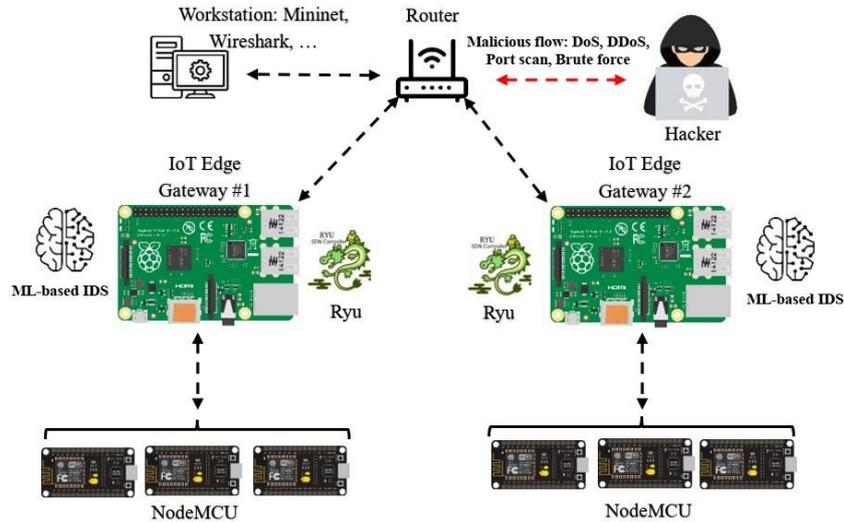

**Figure 1:** IoT-edge testbed topology, illustrating non-SDN and SDN-enabled setups.

To address RQ2, we extended our testbed by integrating the edge gateway with the Ryu controller, establishing an SDN-based environment. Ryu, an open-source Python-based SDN controller [54], provides centralized traffic management, enhancing resource allocation and security analysis. We further utilized Mininet [55] to simulate a realistic SDN infrastructure consisting of eighteen hosts, six switches, and a Ryu controller, mirroring real-world network conditions.

### 4.5. Metrics

We evaluated CPU usage, CPU load, and energy consumption in our test beds in the context of ML-based IDS during cyber threat scenarios. We employed the ANOVA[56] to ensure an objective assessment of the performance of various ML-based IDS.

#### 4.5.1. CPU Load CPU Usage

IDS, especially at the edge and SDN environments. CPU usage measures the percentage of the CPU's current capacity, reflecting how much processing power is dedicated to task execution. High CPU usage in an IDS can signal extensive computational demands, potentially impacting the performance of other tasks and system responsiveness, a concern in resource-limited IoT settings. Efficient IDS, especially those utilizing ML techniques, must manage CPU usage carefully to balance detection accuracy with minimal resource use. Excessive CPU usage can slow IDS's real-time network traffic processing, leading to delays or missed attack detection. On the other hand, CPU load indicates the number of processes waiting to be executed, providing an understanding of the CPU's workload. An increase in CPU load might suggest heavy network traffic or numerous attack attempts, highlighting the risk of system overload. Monitoring CPU load allows for early identification of potential bottlenecks, ensuring that IDS operations do not adversely impact system performance. In SDN-enabled IoT edge systems, adept CPU load management is vital to distribute tasks between IDS and other network efficient functions, ensuring optimal resource allocation and system performance. Both CPU usage and load are pivotal metrics for assessing IDS efficacy in environments where resources are constrained, e.g., at the edge gateway[57][58][59].

#### 4.5.2. CPU Performance Metrics

To assess the computational impact of ML-based IDS, we analyze both CPU load and CPU usage, as these metrics provide complementary insights into system performance. CPU usage is typically expressed as a percentage, indicating the proportion of processing power utilized at a given moment. In contrast, CPU load is presented as a numerical value, representing the average number of active processes waiting for CPU execution over a specific time interval. Moreover, while CPU load can be converted into a percentage, it provides a more detailed view of system stress, especially in multi-core environments. In a multi-core processor, a load value of 1.0 on a single-core system indicates full utilization. In contrast, on a quad-core system, a load of 1.0 suggests that only 25% of the total available processing capacity is used. This distinction is crucial when interpreting our results, as high CPU load does not always imply that the system is at risk of overutilization—it depends on the number of available processing cores and the workload distribution.

### 4.5.3. Energy Consumption

Energy consumption, often measured in watt-hours or joules, quantifies the amount of energy a device or system expended during its operation. In IoT hardware, where many devices are battery-powered or operate in energy-constrained environments, efficient energy consumption is desirable and necessary. Devices (e.g., sensors, actuators) and even more complex IoT nodes must be designed to perform their tasks while consuming minimal energy, ensuring longevity, and reducing the need for frequent battery replacements or recharges. Moreover, IoT devices integrated with SDN bring a new dimension to the energy conversation; SDN centralizes network control, dynamically optimizing network resources based on real-time demands. Although this centralization offers enhanced flexibility and scalability, it also means that the network's core components must be energy efficient. In IoT systems, where potentially thousands or even millions of devices communicate and exchange data, even minor inefficiencies in energy consumption can accumulate, leading to significant energy drains. Integrating ML-based IDS into the edge gateway emphasizes the need to consider energy metrics critically. ML-based IDS are inherently data-intensive, requiring substantial computational resources to process large datasets for detecting and mitigating security threats. Although these systems offer invaluable security enhancements, their operation can be energy-intensive. Therefore, measuring and optimizing the energy consumption of ML-based IDS is crucial to ensure they deliver effective security measures without unduly burdening the system's energy resources. This balance is essential for maintaining the sustainability and efficiency of the edge gateway, where energy efficiency is often a key concern.

We employed PowerTop [60], a robust tool, to precisely gauge and examine the energy consumption in two separate testbed configurations: the edge gateway integrated with SDN and without SDN. PowerTop's sophisticated monitoring capabilities allowed us to gain insights into these testbeds' energy consumption patterns and processor activity.

### 4.5.4. Designed cyber threats

For our research, we focused on analyzing DDoS, DoS, brute force attacks, and the port scan. We chose these specific types of attacks since they were already categorized in the employed dataset. These cyber threats are prevalent and pose substantial risks in the field of cybersecurity. Below, a concise summary of each is presented:

- **A Denial-of-Service (DoS):** At the edge, DoS attacks are critical cybersecurity threats that disrupt device and service operations by flooding systems with excessive requests and consuming vital resources (e.g., bandwidth, processing power, and memory). This overload prevents the system from serving legitimate users, blocking access to essential operations. The distributed, resource-constrained nature of the edge makes them especially susceptible to DoS attacks. The vulnerability of these devices, coupled with their interconnectedness, means that an attack on a single device can significantly compromise the entire network's functionality and security [61].

- **A distributed denial-of-service (DDoS):** A DDoS attack is a coordinated effort where multiple attackers from different locations flood a specific target, such as a server or network at the edge, with excessive traffic. The goal is to deplete the target's resources, causing severe service disruptions or a complete shutdown. Unlike traditional DoS attacks, which come from a single source, DDoS attacks are distributed across numerous sources, making them harder to defend against. This distributed nature makes DDoS attacks especially dangerous at the edge, where the interconnected and resource-constrained devices can exacerbate the attack's impact, potentially crippling the entire network [62].

- **Brute Force:** A brute force attack involves an attacker systematically attempting to gain unauthorized access to a system by trying every possible combination, such as trying every key until one works. With its many interconnected devices and varying security levels, the edge is especially vulnerable to such attacks. Attackers exploit these weaknesses by repeatedly guessing passwords, encryption keys, or access codes, which seriously threatens the integrity and confidentiality of data at the edge gateway[63].

- **Port Scan:** A port scan aims to identify a target system's open ports. By identifying open ports and the services running on them at the edge, attackers can uncover and exploit vulnerabilities, posing a serious threat to the security and integrity of the edge gateway[64].

### 4.5.5. Analysis method for energy consumption, CPU usage, CPU load

We used ANOVA to assess our observed results. ANOVA is an indispensable statistical tool for testing the null hypothesis that posits the equivalence of group means. Our study specifically employed one-way ANOVA to examine the impact of a singular independent variable on the evaluated systems. This method relies on several crucial assumptions, including the necessity for the data to exhibit a normal distribution, the variances between groups being equal (homogeneity of variance), and all observations being independent.

In addition, we conducted 15 separate tests on ML-based IDS to measure CPU load, CPU usage, and energy consumption under various cyber threats. This rigorous approach allowed us to leverage the F statistic, which quantifies the variance ratio between the means of different groups to the variance in the groups. A significant F-statistic, together with a p-value of ≤ 0.05, denotes statistically significant differences between group means, underscoring the efficacy of our testing methodology. By implementing this robust statistical framework, we have thoroughly evaluated the performance of various ML-based IDS models in response to different cyber threats. This analysis has allowed us to identify specific models that demonstrate resilience or efficiency against multiple attacks and require increased computational resources or energy consumption. While CPU load is a key performance metric for IDS evaluation, it is also crucial to consider its impact on IoT device availability and reliability. Excessive CPU consumption by an IDS can degrade the device's primary functions, leading to slow response times or system failures. This is especially critical in real-time applications such as healthcare, industrial automation, and smart home security, where device downtime can have serious consequences. An IDS must enhance security without inadvertently causing an attack such as a DDoS condition due to resource exhaustion. In addition, through these fifteen iterations of testing, ANOVA has enabled us to validate significant differences in IDS performance metrics (e.g., detection accuracy, false positive rates), CPU load, CPU usage, and energy consumption across diverse scenarios. This methodological approach provides a detailed examination of how different IDS models respond to varied threats, establishing a solid statistical foundation for assessing the efficacy of each model in a controlled environment. By distinguishing between performance differences attributable to the models' inherent capabilities and those due to random variation, our use of ANOVA has proven to be critical. It aids in identifying the most resource-efficient and reliable IDS, thereby guiding the selection process for optimal cybersecurity defenses and enhancing our management and understanding of IDS performance under cyber threat conditions [65] [66].

### 4.6. TestSuite

To initiate the research work presented in this paper and to facilitate the environment for further research and testing, we introduce a versatile test suite designed to experiment with and evaluate ML-based IDS in SDN environments. Unlike conventional experimental testbeds, our test suite is an extensible framework equipped with predefined APIs and a selection of pre-integrated algorithms, facilitating the seamless integration and testing of novel IDS models. Another good contribution to our test suite is that users can execute their experiments on it without Raspberry Pi or any other hardware support. As discussed in the previous paragraph, the test suite is developed following the plug-in architecture feature. This ensures that the user can easily integrate their algorithm into the test suite and test the accuracy, energy consumption, and CPU usage with or without security threats. Users can create their own IoT-SDN network and complexity in the network and generate any number of security breaching attacks. This approach not only simplifies the validation process of IDS models in a realistic network scenario but also encourages the exploration of innovative IDS methodologies by providing a solid foundation of tools and benchmarks. We have made the test suite available with the same configuration discussed in Section 4.4. We integrated the same tools for creating an IoT-SDN network, generating security attacks, and measuring IDS accuracy, energy consumption, CPU usage, etc. Through its design, the test suite aims to advance the development and thorough evaluation of cutting-edge IDS solutions, significantly enhancing network security in the era of SDN.

## 5. Experimental Results and Analysis

This section discusses our experimental results and findings. After presenting our results, we conducted an in-depth statistical analysis using ANOVA. This analysis aims to illuminate the implications and insights that emerge from the experimental results, providing an understanding of the efficacy and nuances of each IDS under study.

### 5.1. Experimental finding for RQ1
**CPU Load:**

We tested ML-based IDSs under various cyberattack scenarios to assess their impact and strain on our testbed. The types of cyberattacks we considered include DDoS, DoS, brute force attacks, and the port scan. Moreover, we conducted the ANOVA focusing on CPU load variations in our testbed. Figure 2 illustrates a comparative analysis of the average CPU load among different ML-based IDS models in the presence of various types of cyberattacks. The DL-based IDS (CNN, LSTM, combined model of LSTM and CNN, and EIDM) consistently maintain lower CPU loads across all attack types, demonstrating their efficiency in resource utilization during inference. In contrast, traditional ML-based IDS such as KNN, DT, and RF exhibit significantly higher CPU loads, especially under brute force and DDoS attacks, with KNN and DT being the most resource-intensive. This is because DL models, such as CNN and LSTM, efficiently handle computations in parallel and are optimized for inference. In contrast, traditional models (e.g., KNN and DT) require more repeated, resource-heavy calculations, such as distance computations in KNN or recursive splitting in DTs, especially under large-scale attacks.

**Statistical Findings:**

We conducted an ANOVA, and the results presented in Table 4 illuminate significant differences in CPU load among diverse ML-based IDS under DDoS, underscored by F-statistic of 60.40 and a p-value $< 0.05$. This F-statistic delineates the contrast in CPU load variance across ML-based IDSs against the variance in, highlighting a significant influence of IDS selection on CPU load. The remarkably low p-value corroborates this finding, conclusively demonstrating the substantial differences in CPU load among the IDSs. Furthermore, we observed similar p-values ($< 0.05$) across other attacks, including brute force, DoS, and the port scan, so we do not report them. This reinforces the presence of marked differences in CPU load among diverse ML-based IDS under different cyber threats.

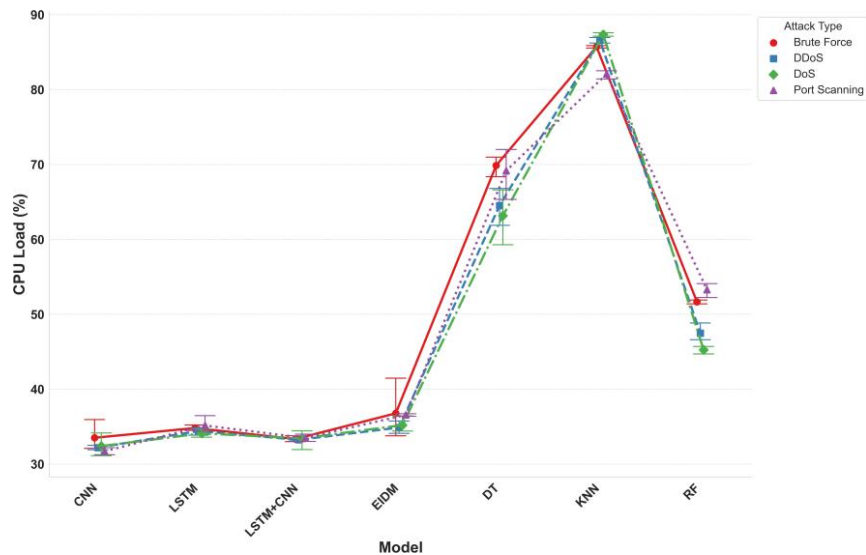

**Figure 2:** The Average CPU load of ML-based IDS under cyber threats.

**Table 4**
ANOVA results: CPU Load for ML-based IDS under DDoS.

| Source | Degrees of Freedom | Sum of Squares | Mean Square | F Statistic | P-value |
|---|---|---|---|---|---|
| Between groups | 6 | 21609.87 | 3601.64 | 60.40 | < 0.05 |
| in groups | 91 | 5426.49 | 59.63 | | |
| Total | 97 | 27036.36 | 278.73 | | |

> **Finding**
>
> DL-based IDS, such as CNN, LSTM, and hybrids, perform more efficiently in managing computational demands across diverse types of cyber threats than traditional ML-based IDS, such as KNN, DT, and RF, as they exhibit higher CPU loads at the edge. This pattern suggests that DL-based IDS' intrinsic efficiency is not attack-specific but rooted in their architecture, making them especially suited for real-time applications at edge gateway.These results are expected, as traditional ML-based IDS (e.g., KNN, DT, RF) perform computationally expensive operations during inference, unlike DL-based IDS, which optimizes processing through parallelization and learned feature extraction.

**CPU Usage:**
Figure 3 compares the average CPU usage of various ML-based IDS models under different cyberattacks. The KNN model consistently exhibits the highest CPU usage across all attack types, indicating its high computational demand, which limits its use in resource-constrained environments. The RF and DT models are also CPU-bound, though they are less intensive than KNN. In contrast, the LSTM model demonstrates the lowest CPU usage, making it the most efficient option for scenarios where minimizing resource consumption is critical. The hybrid of the LSTM and CNN model, along with the CNN and EIDM models, offer a balance between inference accuracy and computational efficiency, making them viable choices for environments with moderate resource availability.

**Statistical Findings:**
Table 5 presents our ANOVA results. Our results reveal significant differences in CPU load among diverse ML-based IDS under DDoS, as evidenced by a compelling F-statistic of 60.39 and a p-value $< 0.05$. This F-statistic highlights the variance in CPU load across IDS groups compared to the variance in, underscoring a significant impact of IDS selection on CPU load. The exceedingly small p-value further supports this conclusion. Moreover, we observed similar p-values (below 0.05) across various cyber threats, such as brute force, DoS, and the port scan, so we do not report those results.

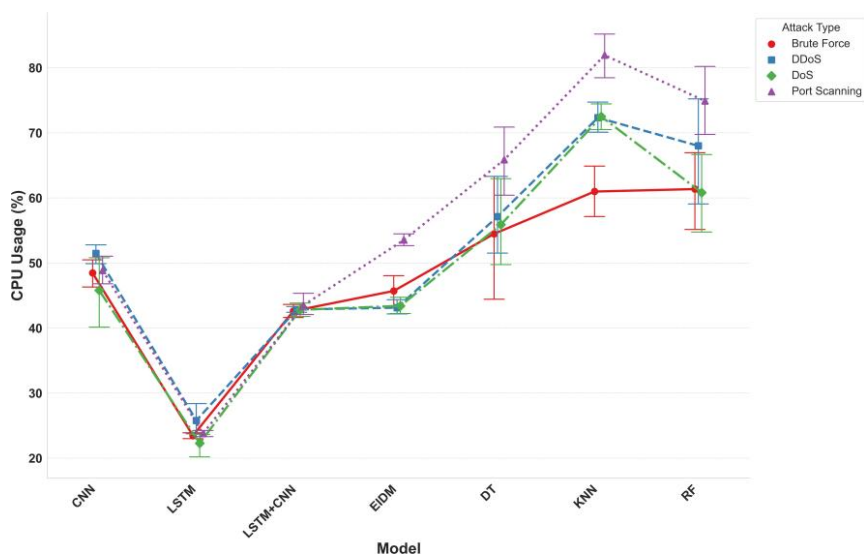

**Figure 3:** The Average CPU usage of ML-based IDS under cyber threats.

**Table 5**
ANOVA results: CPU Usage for ML-based IDS under DDoS.

| Source | Degrees of Freedom | Sum of Squares | Mean Square | F Statistic | P-value |
| --- | --- | --- | --- | --- | --- |
| Between groups | 6 | 21609.86 | 3601.64 | 60.39 | $< 0.05$ |
| in groups | 91 | 5426.49 | 59.62 | | |
| Total | 97 | 27036.36 | 278.73 | | |

> **Finding**
>
> Our analysis reveals that traditional ML-based IDS such as KNN, DT, and RF exhibit increased CPU usage under various cyber threats, thus posing challenges for the edge. Also, LSTM and other DL-based IDS exhibit lower CPU demands. This consistent efficiency across various attacks highlights the benefit of adopting DL-based IDS at the edge gateway. The increased CPU usage of KNN, DT, and RF reflects their reliance on instance-based and tree-splitting operations, which require repeated evaluations. In contrast, DL models efficiently process data in structured layers, reducing computational strain.

**Energy consumption:**
Figure 4 shows that the LSTM and DT models are the most energy-efficient across different types of cyberattacks, consistently exhibiting the lowest energy consumption. The CNN model also performs efficiently, with slightly higher energy usage. The LSTM, CNN model hybrid, and EIDM have moderate energy consumption, balancing complexity and efficiency. In contrast, the KNN model has the highest energy consumption across all scenarios, making it less suitable for energy-constrained environments. The RF model falls in between, with moderate energy demands.

**Statistical Findings:**
We conducted the ANOVA, and the results presented in Table 6 reveal significant differences in energy consumption among diverse ML-based IDS under DDoS, underscored by F-statistic of 57.44 and a p-value of $< 0.05$. This F-statistic delineates the contrast in energy consumption variance across the group of IDSs against the variance in, highlighting a significant influence of IDS selection on energy consumption. The extremely low p-value further supports this conclusion, conclusively demonstrating the substantial differences in energy consumption among the IDSs. In addition, we observed similar p-values ($< 0.05$) for other cyber threats, such as brute force, DoS, and the port scan, so we do not report the results. This observation demonstrates significant differences in energy consumed among various ML-based IDS when faced with differing cyber threats.

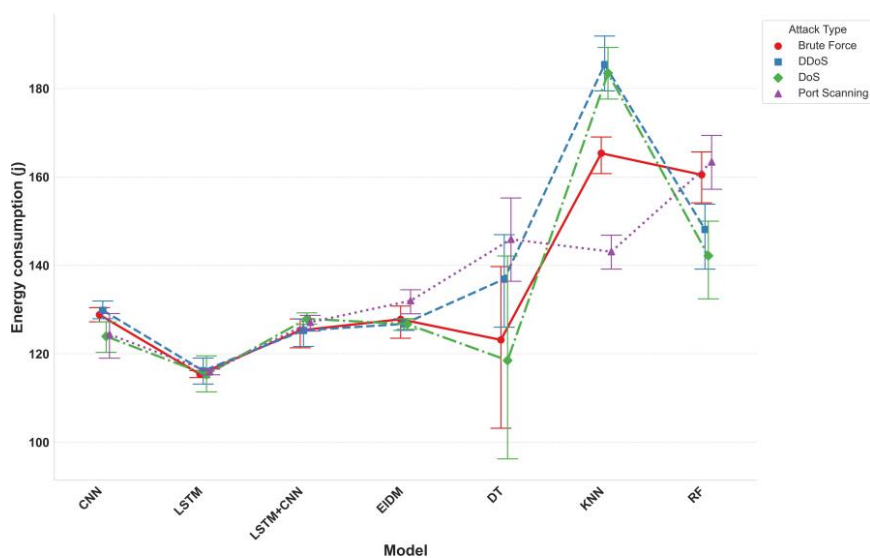

**Figure 4:** The Average Energy consumption of ML-based IDS under cyber threats.

**Table 6**
ANOVA results: energy consumption for ML-based IDS under DDoS.

| Source | Degrees of Freedom | Sum of Squares | Mean Square | F Statistic | P-value |
|---|---|---|---|---|---|
| Between groups | 6 | 47732.07 | 7955.34 | 57.44 | $< 0.05$ |
| in groups | 98 | 13571.72 | 138.48 | | |
| Total | 104 | 61303.80 | 589.45 | | |

> **Finding**
>
> Our analysis concludes a marked discrepancy in energy consumption, with traditional ML-based IDS such as KNN, RF, and DT exhibiting significantly higher energy consumption under cyber threats such as DDoS and brute force, a drawback for energy-constrained at the edge. In contrast, DL-based IDS models, LSTM, CNN, EIDM, and their hybrids excel in energy efficiency, making them the preferable choice for the edge. Traditional ML models' higher energy consumption results from their iterative computations and lack of optimized inference paths, making them less viable for real-time IoT applications where power efficiency is crucial.

### 5.2. Experimental finding for RQ2

This section presents our experimental results for IoT-edge devices with SDN integration during real-time cyber threats.

**CPU Load:**

In Figure 5, we illustrate the CPU load of various ML-based IDS models under different cyberattacks in an SDN-enabled at the edge gateway. The analysis shows that KNN and DT models have the highest CPU load, especially during DDoS and DoS, indicating significant resource demands at the edge. Conversely, the LSTM model demonstrates the lowest CPU load, highlighting its efficiency in resource management. The CNN model also performs efficiently but not as well as LSTM. The LSTM and CNN model hybrid, similar to EIDM, offers balanced performance, making them suitable for scenarios where moderate CPU efficiency is required at the edge.

**Statistical Findings:**

We conducted an ANOVA for the case of the DDoS attack, and the results are presented in Table 7. The results reveal significant differences in CPU load among diverse ML-based IDS under DDoS attack, underscored by an impressive F-statistic of 142.57 and a p-value of $< 0.05$. This F-statistic highlights the variance in CPU load across IDSs compared to the variance in them, indicating a significant impact of IDS selection on CPU load. In addition, consistent p-values ($< 0.05$) were observed across other cyber threats, including brute force, DoS, and the port scan, and we do not report the result. This reinforces the presence of marked differences in CPU load among diverse ML-based IDS when subjected to different cyber threats.

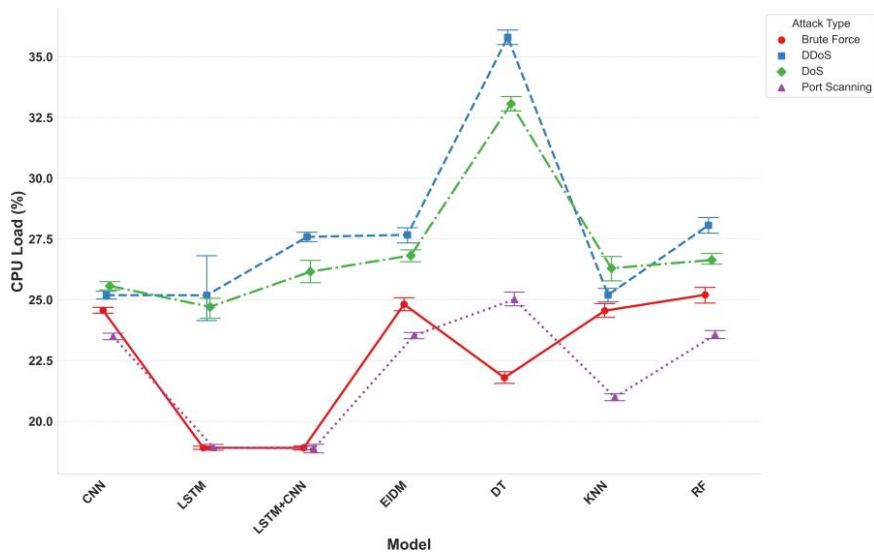

**Figure 5:** The Average CPU load of ML-based IDS under cyber threats.

**Table 7**
ANOVA results: CPU load for ML-based IDS in SDN under DDoS.

| Source | Degrees of Freedom | Sum of Squares | Mean Square | F Statistic | P-value |
|---|---|---|---|---|---|
| Between groups | 6 | 1184.21 | 197.36 | 142.57 | $< 0.05$ |
| in groups | 91 | 125.97 | 1.38 | | |
| Total | 97 | 1310.18 | 13.50 | | |

> **Finding**
>
> The findings demonstrate that traditional ML-based IDS, e.g., DT, exhibit elevated loads under DDoS and DoS. In contrast, DL-based IDSs, including EIDM, LSTM, CNN, and their hybrids, demonstrate superior energy efficiency, making them suitable for SDN-enabled at the edge gateway. The integration of SDN helps balance network resource allocation. Yet, traditional ML-based IDS still exhibit higher CPU load due to their design, reinforcing the efficiency advantage of DL-based models in dynamic network environments.

**CPU Usage:**

Figure 6 shows that CPU usage across various ML-based IDS models in an SDN-enabled edge gateway is fairly consistent across different attack scenarios. Only minor variations are observed, as CNN, LSTM, and hybrid versions demonstrate relatively lower CPU usage, indicating efficient resource management. The DT, KNN, and RF models also show consistent CPU usage across attacks. The EIDM model balances efficiency and performance well.

**Statistical Findings:**

We conducted an ANOVA for the results we got for ML-based IDS in SDN under the DDoS attack. The results presented in Table 8 reveal significant differences in CPU usage among diverse ML-based IDS under DDoS attack, underscored by an impressive F-statistic of 5.94 and a p-value of $< 0.05$. This F-statistic highlights the variance in CPU usage across the group of IDSs compared to the variance in, indicating a significant impact of IDS selection on CPU usage. In addition, we observed a consistently low p-value ($< 0.05$) for other examined cyber threats (not reported in the paper), including brute force, DoS, and port scan, reinforcing the presence of marked differences in CPU usage among diverse ML-based IDS when subjected to different cyber threats.

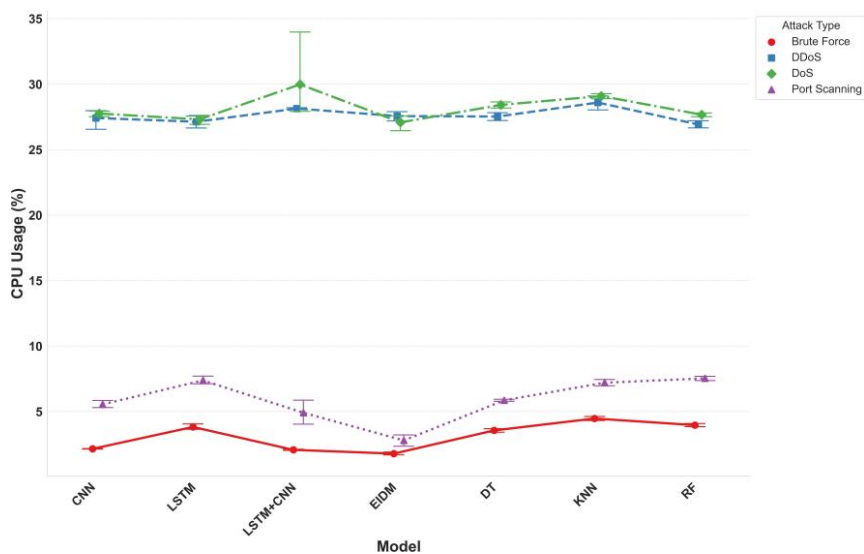

**Figure 6:** The Average CPU usage of ML-based IDS under cyber threats.

**Table 8**
ANOVA results: CPU usage for ML-based IDS in SDN under DDoS.

| Source | Degrees of Freedom | Sum of Squares | Mean Square | F Statistic | P-value |
| --- | --- | --- | --- | --- | --- |
| Between groups | 6 | 27.97 | 4.66 | 5.94 | $< 0.05$ |
| in groups | 91 | 71.32 | 0.78 | | |
| Total | 97 | 99.30 | 1.02 | | |

> **Finding**
>
> In the context of SDN-enhanced IoT, deploying DL-based IDS with advanced models such as CNN, LSTM, EIDM, and their hybrids demonstrates efficient energy consumption. These models achieve reduced CPU usage against brute force and port scan, benefiting from the centralized resource optimization afforded by SDN. Nonetheless, the complexity of DDoS and DoS presents a significant challenge, necessitating increased computational resources. Although SDN optimizes network operations, IDS models such as KNN and RF remain resource-intensive due to their frequent computational overhead. At the same time, DL-based IDS maintains efficiency through batch processing and learned representations.

**Energy consumption:**
Figure 7 depicts the average energy consumption of ML-based IDS models under different attacks in an SDN environment. The results indicate that traditional ML models consume more energy, especially during port scans, e.g., DT, KNN, and RF. In contrast, the EIDM model consistently shows lower energy consumption across all attack types, highlighting its efficiency. The LSTM and CNN models display moderate energy usage, including their hybrid version. Compared to non-SDN environments, the increased energy consumption in the SDN setup is attributed to the SDN controller's active role in traffic management and threat response, which demands more energy resources.

**Statistical Findings:**
We applied ANOVA on energy consumption data across ML-based IDSs in SDN under DDoS. The results, presented in Table 9, reveal significant differences in energy consumption among diverse ML-based IDS under DDoS, underscored by an impressive F-statistic of 18.27 and a p-value of $< 0.05$. This F-statistic highlights the variance in energy consumption across a group of IDSs compared to the variance in, indicating a significant impact of IDS selection on energy consumption. Moreover, a consistently low p-value ($< 0.05$) was observed across other cyber threats, including brute force, DoS, and port scan, so we do not report the results here. This highlights marked differences in CPU usage among diverse ML-based IDS when subjected to examined cyber threats.

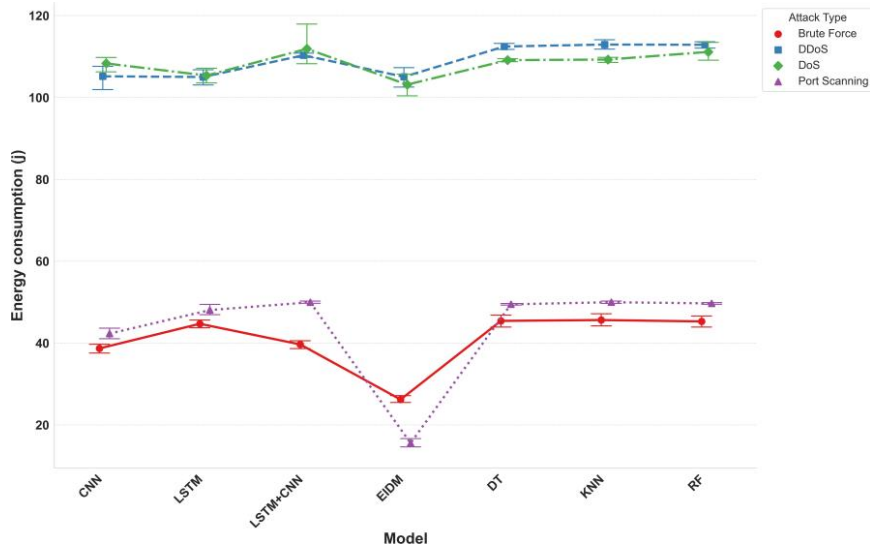

**Figure 7:** The Average Energy consumption of ML-based IDS under cyber threats.

**Table 9**
ANOVA results: Energy consumption for ML-based IDS in SDN under DDoS.

| Source | Degrees of Freedom | Sum of Squares | Mean Square | F Statistic | P-value |
| --- | --- | --- | --- | --- | --- |
| Between groups | 6 | 1263.26 | 210.54 | 18.27 | $< 0.05$ |
| in groups | 91 | 1048.21 | 11.51 | | |
| Total | 97 | 2311.48 | 23.82 | | |

> **Finding**
>
> The findings accentuate the distinct energy efficiency profiles of ML-based IDSs when exposed to various cyber threat scenarios. During brute force and the port scan, traditional ML-based IDS such as DT, KNN, and RF are observed to have higher energy consumption. This indicates that these models are not energy-efficient under the examined conditions due to their complex computational frameworks. On the other hand, DL-based IDS and the EIDM show markedly superior energy efficiency. The reduced energy footprint of DL-based IDS is especially advantageous in the context of the SDN-enabled at the edge, where low energy consumption is crucial due to device constraints and the need for long-term, autonomous operation. The reduction in energy consumption observed in DL-based IDS when integrated with SDN highlights the benefits of centralized network control and optimized workload distribution, making them a more sustainable choice for IoT security.

### 5.3. Analyzing the Impact of SDN on CPU Usage, Load, and Energy Efficiency in ML-Based IDS

Figure 8 demonstrates that integrating SDN with ML-based IDS in the edge gateway significantly improves resource efficiency, reducing energy consumption, CPU usage, and CPU load. The most substantial improvement is in CPU usage, where DL-based IDS, e.g., LSTM and CNN, outperform traditional ML models by efficiently handling complex computations through parallel processing. Additionally, SDN integration reduces CPU load by balancing workloads, essential for real-time threat detection in edge gateway. The observed reduction in energy consumption further highlights the approach's suitability for battery-powered edge gateway, confirming its scalability and practicality for real-world applications.

## 6. ML-Based IDS vs. Signature-Based IDS (Snort)

This section compares our ML-based IDS models and the signature-based Snort IDS to evaluate the performance improvements achieved by leveraging ML-based IDS over traditional detection systems. This comparison is essential to highlight the advantages of ML-based approaches regarding resource efficiency, scalability, and adaptability, especially in edge gateway.

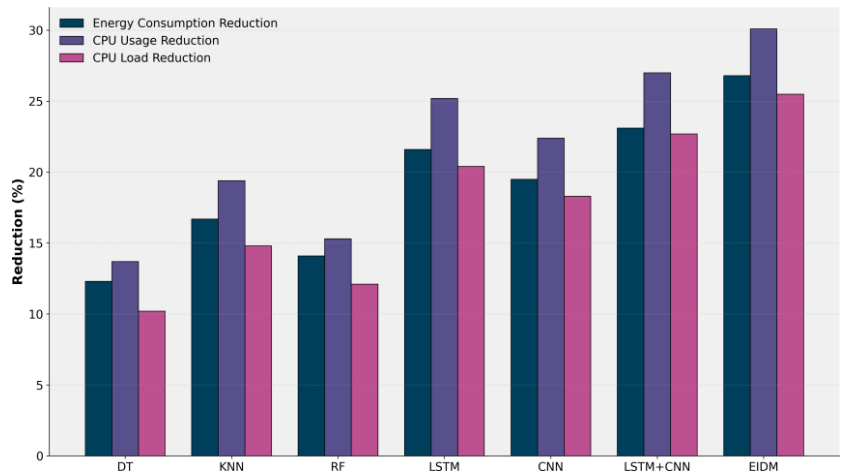

**Figure 8:** Reduction in energy consumption, CPU usage, and CPU load for ML-based IDS models with SDN integration in edge gateway.

The results presented in Table 10 provide a comparative analysis of our ML-based IDS models against the signature-based Snort IDS discussed in other research.
Regarding CPU usage, Snort IDS shows high utilization under heavy traffic due to its reliance on predefined rules and signature matching. In contrast, the ML-based IDS models demonstrate better CPU efficiency. While traditional ML models, e.g., DT and KNN, have higher CPU usage because of iterative computations, DL-based IDS, e.g., LSTM, CNN, and a hybrid of LSTM and CNN, EIDM exhibits lower CPU usage. This is primarily due to DL-based IDS's ability to process data in batches and leverage parallel processing for real-time threat detection. For energy consumption, Table 10 shows that Snort IDS consumes more energy, especially in IoT networks requiring multiple containers. However, our ML-based IDS models, especially DL architectures, e.g., LSTM and EIDM, demonstrate superior energy efficiency. These models optimize resource usage and process data efficiently, making them suitable for resource-constrained edge gateway and highlighting their scalability advantages. Finally, in terms of CPU load, Table 10 indicates that earlier versions of Snort IDS suffer from high CPU load on a single core because of their single-threaded architecture. Although newer versions introduce multi-threading, they still encounter processing bottlenecks under heavy traffic. Conversely, the ML-based IDS models distribute the CPU load more effectively across multiple cores. DL-based IDS, especially LSTM and hybrid architectures, achieve the lowest CPU load levels due to their parallel execution capabilities and efficient handling of sequential data.

## 7. Discussion

Our investigations explored the performance metrics of ML-based IDS with various models, especially in IoT-edge devices with and without SDN integration. Our study was primarily evaluating the impact of these models on CPU load, CPU usage, and energy consumption amidst diverse cyberattack scenarios. The empirical findings revealed significant disparities in resource utilization across different ML-based IDS, shedding light on crucial aspects of their deployment in IoT devices integrated with SDN. The KNN, DT, and RF significantly exhibited higher CPU load, CPU usage, and energy consumption, especially under specific types of cyberattacks. While these models are adept at identifying threats, their resource-intensive nature could pose challenges in the IoT context, where computational resources are often limited. This could lead to diminished performance or instability in environments with constrained resources. Specifically, KNN's higher variance in CPU load and energy consumption, as observed in Tables 4 and 5, stems from its lazy learning approach. Unlike other models, KNN does not build a generalized model during training but instead stores the entire dataset and computes distances at query time. This results in increased processing demands, leading to fluctuations in resource utilization. Such behavior makes KNN less suitable for real-time IDS applications in resource-constrained IoT networks[72] [73]. While CPU load significantly impacts energy consumption, it is not the sole factor. Memory operations, network activity, peripheral devices, and thermal management also contribute to power usage in IoT devices. High data transmission rates and active sensors can increase energy demands, while sustained CPU load may trigger additional energy consumption for cooling mechanisms. Although a strong correlation between CPU load and energy consumption is expected, these factors introduce variations across IDS models. Optimizing IDS efficiency can help balance security and resource constraints in IoT networks. Conversely, the CNN and LSTM models demonstrated greater efficiency in resource utilization. While their architectures are sophisticated and adept at processing complex data structures, they appear to optimize the computational load during inference when employed in IDS. This makes them more suitable for scenarios where resource conservation is critical. However, the complexity of these models introduces its own set of challenges, especially in terms of training and ongoing maintenance in the dynamic landscape of IoT devices integrated with SDN.

**Table 17**
Comparative Resource Utilization of ML-Based IDS and Snort IDS Based

| Metric | Snort IDS | ML-Based IDS (Our Findings) |
|---|---|---|
| **CPU Usage** | - **High Traffic Conditions:** CPU usage can reach its maximum during initialization with many active rules [67].<br>- **Multi-Core Systems:** Snort 3.0 utilizes a significant portion of CPU resources on a multi-core processor [68] [69]. | - **Traditional ML Models (DT, KNN, RF):** Tend to exhibit higher CPU usage during real-time cyber threats, especially those requiring intensive computations.<br>- **DL-Based Models (CNN, LSTM, Hybrid of LSTM and CNN and EIDM):** Show lower CPU usage compared to traditional ML models, with LSTM models demonstrating the most efficient utilization due to sequential data processing and parallelization. |
| **Energy Consumption** | - **IoT Deployment:** Deployment of Snort on IoT gateways results in considerable energy consumption [70]. | - **Traditional ML-based IDS:** Generally consume more energy during inference cycles due to repetitive computations.<br>- **DL-Based Models:** Exhibit better energy efficiency, especially models that combine convolutional and sequential layers, benefiting from optimized processing structures. |
| **CPU Load** | - **Single-Core Utilization:** Older Snort versions (pre-3.0) lead to high load on a single core under heavy traffic [71].<br>- **Multi-Core Systems:** Updated versions distribute the load but still face processing bottlenecks under extensive traffic [71]. | - **Traditional ML-based IDS:** Often show higher CPU load during complex attack scenarios.<br>- **DL-Based Models:** Maintain a lower CPU load, benefiting from parallel processing capabilities, with hybrid models showing the most balanced load distribution. |

The balance between detection efficiency and resource consumption is especially critical at edge gateway, where devices often have limited processing power and energy reserves. This balance is closely tied to several United Nations Sustainable Development Goals (SDGs), especially SDG 9 (Industry, Innovation, and Infrastructure), SDG 11 (Sustainable Cities and Communities), and SDG 13 (Climate Action). Optimizing IDS deployment in smart cities strengthens cybersecurity infrastructure, directly supporting SDG 9 while fostering resilient, sustainable urban environments in line with SDG 11. Furthermore, by prioritizing energy-efficient IDS solutions, this research contributes to SDG 13, promoting responsible resource consumption and mitigating the environmental impact of growing IoT networks [74].

To aid IoT developers in selecting appropriate IDS solutions, we provide detailed guidelines in Table 11 and Table 12, outlining the performance trade-offs of seven different ML-based IDS models for IoT devices examined in this paper, both with and without SDN integration. These insights enable developers to make informed decisions, ensuring the optimal balance between security and resource efficiency during application development. We use graphical indicators (smiley faces) instead of numerical values to provide an intuitive, high-level comparison of IDS performance. This visual approach simplifies decision-making for IoT developers, aligning with similar methodologies used in prior work [75]. Moreover, all corresponding numerical values related to CPU usage, CPU load, and energy consumption are presented in the Figures and Tables in Section 5.

On the other hand, to the best of our knowledge, only Tekin et al. [12] have explored a similar direction in evaluating the performance of ML-based IDS in IoT systems. However, our study takes a fundamentally different approach, especially in how computational resources are classified and utilized, which plays a critical role in the effectiveness and scalability of IoT systems. While Tekin et al. focus on energy consumption and inference times using Raspberry Pi as an IoT device, our study emphasizes the advantages of processing data at the edge, especially regarding energy efficiency, CPU load, and usage. We show how models such as DT and RF benefit from edge processing, reducing latency and improving responsiveness, especially when combined with SDN, which optimizes network traffic and resource allocation. Our findings underscore the importance of balancing computational tasks across the network using SDN to maintain performance, unlike Tekin et al. [12], who do not explore the impact of edge computing or SDN integration.

## 8. Threat and validity

Empirical research inevitably encounters issues related to the validity of findings. In light of this, the present section seeks to identify and discuss possible threats to our research's validity, per the recommendations of Wohlin et al. [76].

**Table 11**
Guideline for selecting seven ML-based IDS in edge gateway.

| Metric | DT | KNN | RF | CNN | LSTM | CNN | LSTM+CNN | EIDM |
|---|---|---|---|---|---|---|---|---|
| CPU load | ☹ | ☹ | 😐 | 🙂 | 🙂 | 🙂 | 🙂 | 🙂 |
| CPU usage | ☹ | ☹ | ☹ | 🙂 | 🙂 | 🙂 | 🙂 | 🙂 |
| Energy consumption | 😐 | ☹ | ☹ | 🙂 | 🙂 | 🙂 | 🙂 | 🙂 |

**Table 12**
Guideline for selecting seven ML-based IDS in SDN-edge gateway.

| Metric | DT | KNN | RF | CNN | LSTM | CNN | LSTM+CNN | EIDM |
|---|---|---|---|---|---|---|---|---|
| CPU load | ☹ | ☹ | ☹ | 🙂 | 🙂 | 🙂 | 🙂 | 🙂 |
| CPU usage | 😐 | 😐 | 😐 | 😐 | 😐 | 😐 | 😐 | 😐 |
| Energy consumption | 😐 | 😐 | 😐 | 😐 | 😐 | 😐 | 😐 | 😐 |

The energy consumption and CPU usage in all ML-based IDS lowered during the brute force attack and port scan.

### 8.1. Internal Threats

During our empirical study on ML-based IDS in the context of IoT devices with IoT devices integrated with SDN, we recognized the existence of internal obstacles that impact the credibility of our findings. The precision of our performance measures is of utmost importance, namely the measurement of CPU load, CPU usage, and energy consumption in these intricate network settings. The complex characteristics of IoT devices and the adaptable structure of SDN provide significant difficulties in guaranteeing accurate and dependable performance evaluations. To address these concerns, we performed fifteen experiments on our testbeds. To improve the trustworthiness of our results in the context of SDN and IoT, we utilized average values to reduce the impact of network or hardware differences and ambient factors. In addition, the cyber threat simulations were conducted using highly practiced cyber security testing mechanisms in academic research and industries in IoT-edge devices integrated with SDN. This work aims to tackle internal risks associated with the setup and precision of ML-based IDS, improving their usefulness and significance in these fast-advancing technical fields.

### 8.2. External Threats:

The landscape of network security, especially in IoT-edge devices and IoT-edge devices integrated with SDN realms, is increasingly challenged by external threats. These range from sophisticated cyberattacks such as DoS, DDoS, and brute force attacks to more subtle, yet equally harmful, reconnaissance methods such as a port scan. These threats highlight the urgent need for robust and adaptable IDS solutions. Integrating ML into IDS presents promising advancements in threat detection and mitigation. However, this integration faces challenges due to the complexity of IoT-edge devices, which are marked by numerous interconnected devices, and the dynamic nature of SDN architectures. IDS solutions must be precise in threat detection while also being resource-efficient. Our research evaluates ML-based IDS based on CPU usage, CPU load, and energy consumption, especially under real-time cyber threats. These metrics are vital to ensure that ML-based IDS are effective in protecting networks against external threats and sustainable in their operation. They help maintain a crucial balance between security and performance in the complex ecosystems of IoT devices and IoT devices integrated with SDN. Additionally, to ensure the transparency and reproducibility of our study, we have provided detailed information about the experimental setup and made our testbed and results publicly available for further research [77]. By adopting these measures, we have attempted to provide robust validation and increase the inability to reject our findings among practitioners and researchers.

## 9. Conclusion

This paper presents a comparative analysis of the ML-based IDS in IoT-edge devices and IoT-edge devices integrated with SDN under different cyberattack scenarios, resulting in comprehension. In IoT systems, conventional ML models (e.g., KNN and DT) often experience increased CPU load and CPU usage, especially when subjected to DoS and DDoS cyber threats. This suggests that these models have limits in resource-limited situations. In contrast, DL-based IDS (e.g., CNN and LSTM) exhibit reduced CPU usage, indicating improved efficiency and compatibility with IoT security. A consistent energy consumption pattern was identified across attack types in both scenarios, encompassing advanced neural networks and conventional methods. The consistent energy efficiency of these models, independent of their computing complexity, highlights their efficacy and long-term viability for use in different network environments. The findings emphasize the significance of choosing ML-based IDS according to their computational efficiency and energy consumption to achieve optimal performance in networks with limited resources. It is imperative to thoroughly evaluate the scalability and robustness of ML-based IDS in future research, especially in more significant and more complex network environments. This assessment will explain their ability to adjust to changing cyber threats. Furthermore, it is crucial to evaluate the influence of new technologies, e.g., 5G and edge computing, on the efficacy and suitability of ML-based IDS in advanced network infrastructures.

Future research directions should pivot towards optimizing ML-based IDS for enhanced scalability, real-time processing, and energy consumption. The overarching challenge is to develop effective threat detection models that minimally impact system resources. Furthermore, integrating these models into existing IoT devices and IoT devices integrated with SDN infrastructures presents additional challenges, including ensuring compatibility, scalability, and ease of maintenance.

## A. Conflict of interest

The authors declare that they have no known conflict of interest or personal relationships that could have appeared to influence the work reported in this paper.

## B. Acknowledgement

The authors thank Dr. Karim A. Emara et al. for collaborating to share the EIDM-IDS source code.

# Appendix

Table 13: Abbreviations used in this research.

| Abbreviation | Meaning |
| --- | --- |
| AI | Artificial Intelligence |
| ANOVA | Analysis of Variance |
| ANN | Artificial Neural Network |
| BT | Boosting Tree |
| CPU | Central Processing Unit |
| DAE | Deep Autoencoder |
| DDoS | Distributed Denial-of-Service |
| DL | Deep Learning |
| DoS | Denial-of-Service |
| DT | Decision Tree |
| GPU | Graphics Processing Unit |
| IDS | Intrusion Detection System |
| IoT | Internet of Things |
| KNN | K-Nearest Neighbor |
| LR | Logistic Regression |
| LSTM | Long Short-Term Memory |
| CNN | Convolutional Neural Network |
| MCU | Microcontroller Unit |
| MITM | Man-in-the-Middle |
| ML | Machine Learning |
| MTD | Moving Target Defense |
| NB | Naïve Bayes |
| R2L | Root to Local |
| RF | Random Forest |
| RNN | Recurrent Neural Network |
| SDN | Software-Defined Networking |
| SDPN | Stacked-Deep Polynomial Network |
| SMO | Spider Monkey Optimization |
| SMOTE | Synthetic Minority Oversampling Technique |
| SNN | Spiking Neural Network |
| SVM | Support Vector Machine |
| U2R | User to Root |
| WFEU | Wrapper Feature Extraction Unit |
| WSN | Wireless Sensor Network |